\documentclass[aps,prl,twocolumn,showpacs,preprintnumbers,amsmath,amssymb,superscriptaddress]{revtex4}%

\usepackage{graphicx}% Include figure files
\usepackage{dcolumn}% Align table columns on decimal point
\usepackage{bm}% bold math
\usepackage{color}

\begin{document}

%\preprint{PREPRINT (\today)}
%
%*******************************************
\title{ Magnetic field dependence of the oxygen isotope effect on the
magnetic penetration depth in hole-doped cuprate superconductors}
\author{R.~Khasanov}
\email{rustem.khasanov@physik.unizh.ch}
 \affiliation{Physik-Institut der Universit\"{a}t
Z\"{u}rich, Winterthurerstrasse 190, CH-8057 Z\"urich,
Switzerland}
\author{A.~Shengelaya}
 \affiliation{Physik-Institut der Universit\"{a}t
Z\"{u}rich, Winterthurerstrasse 190, CH-8057 Z\"urich,
Switzerland}
 \affiliation{Physics Institute of Tbilisi State
University, Chavchavadze 3, GE-0128 Tbilisi, Georgia }
\author{D. Di~Castro}
\affiliation{Physik-Institut der Universit\"{a}t Z\"{u}rich,
Winterthurerstrasse 190, CH-8057 Z\"urich, Switzerland}
\affiliation{INFM-Coherentia and Dipartimento di Fisica,
Universita' di Roma "La Sapienza", P.le A. Moro 2, I-00185 Roma,
Italy}
\author{D.G.~Eshchenko}
 \affiliation{Physik-Institut der Universit\"{a}t
Z\"{u}rich, Winterthurerstrasse 190, CH-8057 Z\"urich,
Switzerland}
 \affiliation{Laboratory for Muon Spin Spectroscopy,
Paul Scherrer Institut, CH-5232 Villigen PSI, Switzerland}
\author{I.M.~Savi\'c }
\affiliation{Faculty of Physics, University of Belgrade, 11001
Belgrade, Serbia and Montenegro}
\author{K.~Conder}
\affiliation{ Laboratory for Neutron Scattering, ETH Z\"urich and
Paul Scherrer Institut, CH-5232 Villigen PSI, Switzerland}
\author{E.~Pomjakushina}
\affiliation{ Laboratory for Neutron Scattering, ETH Z\"urich and
Paul Scherrer Institut, CH-5232 Villigen PSI, Switzerland}
\author{J.~Karpinski}
\affiliation{Solid State Physics Laboratory, ETH 8093 Z\"urich,
Switzerland}
\author{S.~Kazakov}
\affiliation{Solid State Physics Laboratory, ETH 8093 Z\"urich,
Switzerland}
\author{H.~Keller}
\affiliation{Physik-Institut der Universit\"{a}t Z\"{u}rich,
Winterthurerstrasse 190, CH-8057 Z\"urich, Switzerland}
%

%now the abstract***************************
\begin{abstract}
%\widetext
%\narrowtext

The magnetic field dependence of the oxygen-isotope
($^{16}$O/$^{18}$O) effect (OIE) on the in-plane magnetic field
penetration depth $\lambda_{ab}$ was studied in the hole-doped
high-temperature cuprate superconductors YBa$_2$Cu$_4$O$_8$,
Y$_{0.8}$Pr$_{0.2}$Ba$_2$Cu$_3$O$_{7-\delta}$, and
Y$_{0.7}$Pr$_{0.3}$Ba$_2$Cu$_3$O$_{7-\delta}$. It was found that
$\lambda_{ab}$ for the $^{16}$O substituted samples increases
stronger with increasing magnetic field than for the $^{18}$O
ones. The OIE on $\lambda_{ab}$ decreases by more than a factor of
two with increasing magnetic field from $\mu_0H=0.2$~T to
$\mu_0H=0.6$~T. This effect can be explained by the isotope
dependence of the in-plane charge carrier mass $m^\ast_{ab}$.

\end{abstract}
%*******************************************
%~\\
\maketitle

%Introduction

The presence of nodes in the superconducting gap is probably one of
the most significant features of high-temperature cuprate
superconductors (HTS) which has attracted considerable theoretical
and experimental attention in recent years (see e.g.
\cite{Scalapino95,Tsuei94}). The fourfold symmetric nature of the
$d$-wave order parameter, together with the presence of gap nodes on
the Fermi surface, open possibilities for novel effects to be
observable in HTS. One of the most remarkable effect is the magnetic
field dependence of the in-plane magnetic penetration depth
$\lambda_{ab}$ observed in various HTS in the mixed state (see e.g.
\cite{Sonier00} and references therein).
It was shown that the field dependent correction to $\lambda_{ab}$
arises from the nonlocal and nonlinear response of a
superconductor to an applied magnetic field \cite{Amin00,Amin99}.
In contrast to conventional extreme type-II BCS superconductors,
both nonlinear and nonlocal corrections to $\lambda_{ab}$ were
found to be very strong in HTS \cite{Sonier00,Amin00,Amin99}.

The nonlinear correction to $\lambda_{ab}$  arises from magnetic
field induced quasiparticle excitation in the gap nodes
\cite{Volovik93}. The excitation energy $\varepsilon$ is associated
with the pair momentum $2{\bf q}=m^\ast {\bf v}_s$ (${\bf v}_s$ is
the local superfluid velocity, and $m^\ast$ is the charge carrier
mass) and the Fermi velocity ${\bf v}_F$~\cite{Won01}:
\begin{equation}
\varepsilon = {\bf v}_F\cdot{\bf q}.
 \label{eq:excitation_energy}
\end{equation}
The density of the delocalized states was found to increase
proportionally to $\sqrt{H}$ \cite{Volovik93}. The nonlocal
correction to $\lambda$ appears from the response of electrons with
momenta on the Fermi surface close to the gap nodes. This is because
the coherence length $\xi$, being inversely proportional to the gap,
becomes very large close to the nodes and formally diverges at the
nodal points. Thus, there exist areas on the Fermi surface where
$\lambda/\xi\lesssim 1$, and the response of a superconductor to an
applied magnetic field becomes nonlocal \cite{Amin99}.

A simple analysis reveals that both the nonlinear and the nonlocal
corrections depend on the supercarrier mass $m^\ast$. Indeed, the
nonlinear correction to $\lambda_{ab}$ depends on the excitation
energy $\varepsilon$ [see Eq.~(\ref{eq:excitation_energy})] and,
therefore, will increase with increasing $\varepsilon$. Bearing in
mind that $|{\bf v}_F|\propto 1/m^\ast$ and $|{\bf v}_s|\propto
1/m^\ast$ \cite{Volovik93}, the excitation energy is proportional to
$1/m^\ast$. Thus the nonlinear correction to $\lambda_{ab}$
decreases with increasing supercarrier mass $m^\ast$. On the other
hand, the nonlocal correction depends on the Fermi surface area
where, due to the growing of $\xi$ close to the nodes,
$\lambda/\xi\lesssim 1$ \cite{Amin99}. Therefore, the nonlocal
correction will decrease with increasing Ginzburg-Landau parameter
$\kappa=\lambda_{ab}/\xi_0$. Taking into account that within the
simple London model $\lambda^{-2}\propto 1/m^\ast$ and $\xi_0\propto
|{\bf v}_F|/\Delta_0$, one can easily get $\kappa\propto
(m^\ast)^{3/2}$ ($\xi_0$ and $\Delta_0$ are the the mean values of
the coherence length and the superconducting gap, respectively).
This implies that the nonlocal correction to $\lambda_{ab}$, in
full analogy with the nonlinear one, decreases with increasing
$m^\ast$. As a consequence, the magnetic field dependence of
$\lambda_{ab}$ is expected to be stronger for a superconductor
with a smaller $m^\ast$.

\begin{figure*}[htb]
%\centering
\includegraphics[width=1.0\linewidth]{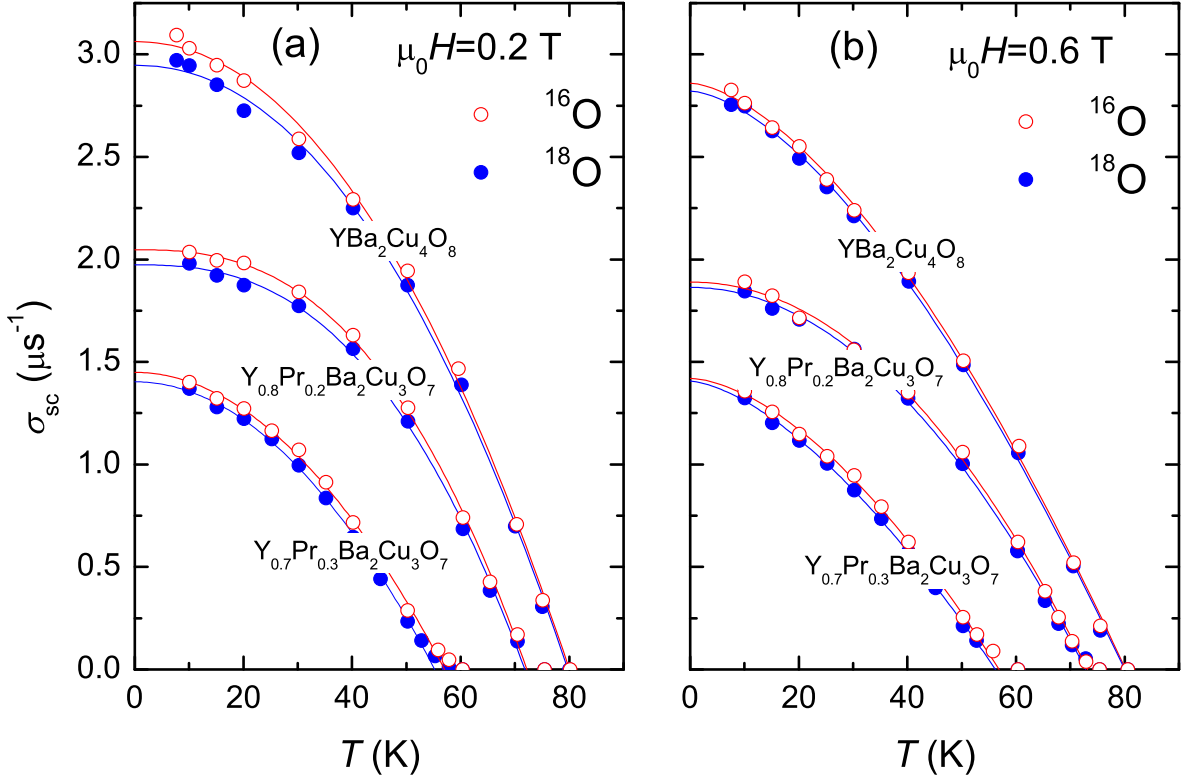}
%\vspace{-0.8cm}
%
\caption{(Color online) Temperature dependences of the
superconducting part of the $\mu$SR depolarization rate
$\sigma_{sc}$ of $^{16}$O ($\circ$) and $^{18}$O ($\bullet$)
substituted YBa$_2$Cu$_4$O$_8$,
Y$_{0.8}$Pr$_{0.2}$Ba$_2$Cu$_3$O$_{7-\delta}$, and
Y$_{0.7}$Pr$_{0.3}$Ba$_2$Cu$_3$O$_{7-\delta}$. (a) --
$\mu_0H=0.2$~T. (b) -- $\mu_0H=$0.6~T. The errors are smaller than
the size of the data points. The solid lines correspond to fits to
the power law $\sigma_{sc}(T)/\sigma_{sc}(0)=1- (T/T_{c})^n$ with
parameters listed in the Table~\ref{Table:OIE-results}.}
 \label{figure:sigma}
\end{figure*}

A study of the field dependence of the isotope effect on
$\lambda_{ab}$ can serve as a direct test of this prediction.
Indeed, several oxygen isotope ($^{16}$O/$^{18}$O) effect (OIE)
studies of the in-plane penetration depth $\lambda_{ab}$ in
various cuprate families
(Y$_{1-x}$Pr$_x$Ba$_2$Cu$_3$O$_{7-\delta}$
\cite{Zhao95,Khasanov03,Khasanov03a,Khasanov04,Khasanov04a,Khasanov06},
YBa$_2$Cu$_4$O$_8$ \cite{Khasanov06}, La$_{2-x}$Sr$_{x}$CuO$_{4}$
\cite{Zhao97,Zhao98,Hofer00,Khasanov04a, Khasanov06}, and
Bi$_{1.6}$Pb$_{0.4}$Sr$_{2}$Ca$_{2}$Cu$_{3}$O$_{10+\delta}$~
\cite{Zhao01}) showed  a pronounced oxygen-mass dependence of the
in-plane supercarrier mass with $^{18}m_{ab}^\ast>\
^{16}m_{ab}^\ast$. This implies that $^{16}\lambda_{ab}$ should
increase stronger with magnetic field than $^{18}\lambda_{ab}$.
Bearing in mind that $^{18}\lambda_{ab}$ is always larger than
$^{16}\lambda_{ab}$
\cite{Zhao95,Khasanov03,Khasanov03a,Khasanov04,Khasanov04a,
Khasanov06,Zhao97,Zhao98,Hofer00,Zhao01,Tallon05} the OIE on
$\lambda_{ab}$
[$\Delta\lambda_{ab}/\lambda_{ab}=($$^{18}\lambda_{ab}-
$$^{16}\lambda_{ab})/$$^{16}\lambda_{ab}$] should decrease with
increasing magnetic field (hereafter the indices 16 and 18 denote
$^{16}$O and $^{18}$O substituted samples, respectively).

Here, we report a study of the magnetic field dependence of the
OIE on $\lambda_{ab}$ in YBa$_2$Cu$_4$O$_8$ and
Y$_{1-x}$Pr$_{x}$Ba$_2$Cu$_3$O$_{7-\delta}$ ($x=0.2$, 0.3) by
means of the muon-spin rotation ($\mu$SR) technique. The isotope
shift of $\lambda_{ab}$ at $T=0$  was found to decrease from
$\Delta\lambda_{ab}/\lambda_{ab}=2.2(6)$\%, 2.4(4)\%, and 2.1(6)\%
at $\mu_0H=0.2$~T to 0.9(6)\%, 1.0(4)\%, and 0.9(6)\% at
$\mu_0H=0.6$~T for YBa$_2$Cu$_4$O$_8$,
Y$_{0.8}$Pr$_{0.2}$Ba$_2$Cu$_3$O$_{7-\delta}$, and
Y$_{0.7}$Pr$_{0.3}$Ba$_2$Cu$_3$O$_{7-\delta}$, respectively.
Moreover, measurements of the magnetic field dependence of
$\lambda_{ab}$ at $T=20$~K for YBa$_2$Cu$_4$O$_8$ reveal that for
$\mu_0H\gtrsim0.2$~T, the penetration depth $\lambda_{ab}$
increases stronger for the $^{16}$O substituted sample than for
the $^{18}$O substituted one, consistent with the finding
$^{18}m^\ast_{ab}>^{16}m^\ast_{ab}$.

%Sample preparation and experimental techniques

Details on the sample preparation for YBa$_2$Cu$_4$O$_{8}$ and
Y$_{1-x}$Pr$_{x}$Ba$_2$Cu$_3$O$_{7-\delta}$ ($x=0.2$, 0.3) can be
found elsewhere \cite{Karpinski89,Conder01}. Oxygen isotope
exchange was performed during heating the samples in $^{18}$O$_2$
gas. In order to ensure that the $^{16}$O and $^{18}$O substituted
samples are the subject of the same thermal history, the annealing
of the two samples was performed simultaneously in $^{16}$O$_2$
and $^{18}$O$_2$ (95\% enriched ) gas, respectively
\cite{Conder01}. The $^{18}$O content in all the samples, as
determined from a change of the sample weight after the isotope
exchange, was found to be 82(2)\%.

The transverse-field $\mu$SR experiments were performed at the Paul
Scherrer Institute (PSI), Switzerland, using the $\pi$M3 $\mu$SR
facility. In a powder sample the magnetic penetration depth
$\lambda$ can be extracted from the muon-spin depolarization rate
$\sigma(T) \propto 1/\lambda^{2}(T)$, which probes the second moment
$\langle \Delta B^{2}\rangle$ of the probability distribution of the
local magnetic field function $p(B)$ in the mixed state
\cite{Zimmermann95}. For highly anisotropic layered superconductors
(like cuprate superconductors) $\lambda$ is mainly determined by the
in-plane penetration depth $\lambda_{ab}$ \cite{Zimmermann95}: $
\sigma(T) \propto 1/\lambda_{ab}^{2}(T)$.
The depolarization rate $\sigma$ was extracted from the $\mu$SR
time spectra using a Gaussian relaxation function $R(t) =
\exp[-\sigma^{2}t^{2}/2]$. The superconducting contribution
$\sigma_{sc}$ was then obtained by subtracting the dipolar
contribution $\sigma_{nm}$ measured above $T_{c}$ as
$\sigma_{sc}^2=\sigma^2-\sigma_{nm}^2$.

%Results and discussion

Fig.~\ref{figure:sigma} shows the temperature dependences of
$\sigma_{sc}$ for the samples studied in this work, measured after
field-cooling the samples from far above $T_c$ in $\mu_0H$=0.2~T (a)
and 0.6~T (b). In both fields $\sigma_{sc}$ for the $^{18}$O
substituted samples is systematically lower than those for the
$^{16}$O substituted samples. The data in Fig.~\ref{figure:sigma}
were fitted with the power law $\sigma_{sc}(T)/\sigma_{sc}(0)=1-
(T/T_{c})^n$ \cite{Zimmermann95} with $\sigma_{sc}(0)$, $n$, and
$T_c$ as free parameters. The values of $T_c$ and $\sigma_{sc}(0)$,
obtained from the fits are listed in Table~\ref{Table:OIE-results}.
From $\sigma_{sc}(0)$ the relative isotope shift of the in-plane
penetration depth $\Delta \lambda_{ab}(0)/\lambda_{ab}(0) =
-0.5\cdot [^{18}\sigma_{sc}(0) -$$ ^{16}\sigma_{sc}(0)] /
^{16}\sigma_{sc}(0)$ was determined (see
Table~\ref{Table:OIE-results}). The values of $\Delta
\lambda_{ab}(0)/\lambda_{ab}(0)$ were corrected for the incomplete
$^{18}$O content (82\%) of the $^{18}$O substituted samples.

\begin{table*}
\caption[~]{\label{Table:OIE-results}Summary of the OIE results
for YBa$_2$Cu$_4$O$_8$,
Y$_{0.8}$Pr$_{0.2}$Ba$_2$Cu$_3$O$_{7-\delta}$, and
Y$_{0.7}$Pr$_{0.3}$Ba$_2$Cu$_3$O$_{7-\delta}$ extracted from the
experimental data (see text for an explanation). The values of
$\Delta\lambda_{ab}(0,H)/\lambda_{ab}(0,H)$ were corrected for incomplete $^{18}$O exchange of 82\%.} %
%\begin{center}
\begin{tabular}{lccccc|cccccccc} \hline\hline
&\multicolumn{5}{c}{$\mu_0H=0.2$~T}&\multicolumn{5}{c}{$\mu_0H=0.6$~T}\\
 %\hline
Sample&$^{16}T_c$&$^{16}\sigma_{sc}(0)$&$^{18}T_c$&$^{18}\sigma_{sc}(0)$&$\frac{\Delta\lambda_{ab}(0)}{\lambda_{ab}(0)}$&
       $^{16}T_c$&$^{16}\sigma_{sc}(0)$&$^{18}T_c$&$^{18}\sigma_{sc}(0)$&$\frac{\Delta\lambda_{ab}(0)}{\lambda_{ab}(0)}$\\
&[K]&[$\mu$s$^{-1}$]&[K]&[$\mu$s$^{-1}$]&[\%]&[K]&[$\mu$s$^{-1}$]&[K]&[$\mu$s$^{-1}$]&[\%]\\
 \hline
YBa$_2$Cu$_4$O$_8$
                                              &80.03(45)&3.06(2)&79.77(45)&2.95(2)&2.2(6)&
                                              80.17(37)&2.86(2)&79.99(37)&2.82(2)&0.9(6)\\
Y$_{0.8}$Pr$_{0.2}$Ba$_2$Cu$_3$O$_{7-\delta}$&72.22(38)&2.05(1)&71.82(44)&1.97(1)&2.4(4)&
                                              73.19(35)&1.89(1)&72.84(34)&1.86(1)& 1.0(4)\\
Y$_{0.7}$Pr$_{0.3}$Ba$_2$Cu$_3$O$_{7-\delta}$&57.05(57)&1.45(1)&55.23(34)&1.40(1)&2.1(6)&
                                              56.88(39)&1.42(1)&56.21(31)&1.40(1)&0.9(6) \\
 \hline \hline
\end{tabular}

\end{table*}

Two important points should be considered:
{(i)} For  all the samples (except $^{18}$O substituted
Y$_{0.7}$Pr$_{0.3}$Ba$_2$Cu$_2$O$_{7-\delta}$) $\sigma_{sc}(0)$
decreases with increasing magnetic field. The changes of
$\sigma_{sc}(0)$ for both oxygen isotopes were found to be
$[\sigma_{sc}(0.6{\rm T})-\sigma_{sc}(0.2{\rm
T})]/\sigma_{sc}(0.2{\rm T})\simeq-5$\%, $\simeq-6$\%, and
$\simeq-1$\% for YBa$_2$Cu$_4$O$_8$,
Y$_{0.8}$Pr$_{0.2}$Ba$_2$Cu$_3$O$_{7-\delta}$, and
Y$_{0.7}$Pr$_{0.3}$Ba$_2$Cu$_3$O$_{7-\delta}$, respectively. Note,
that the values for YBa$_2$Cu$_4$O$_8$ and
Y$_{0.8}$Pr$_{0.2}$Ba$_2$Cu$_3$O$_{7-\delta}$ are in fair agreement
with those reported in the literature for optimally doped
YBa$_2$Cu$_3$O$_{7-\delta}$  and La$_{1.85}$Sr$_{0.15}$CuO$_4$
\cite{Sonier00}. The value of $-1$\% obtained for
Y$_{0.7}$Pr$_{0.3}$Ba$_2$Cu$_3$O$_{7-\delta}$ is somehow small. The
reason for that is probably the enhancement of the magnetic
contribution to the muon relaxation rate $\sigma$ at higher fields,
observed in similar samples below $\sim10$~K \cite{Khasanov03}.
{(ii)} In both fields $^{16}\sigma_{sc}(0)>$$^{18}\sigma_{sc}(0)$
[$^{16}\lambda_{ab}(0)<$$^{18}\lambda_{ab}(0)$], implying that the
in-plane charge carriers are heavier in the $^{18}$O substituted
samples than in the $^{16}$O substituted ones, as it was previously
observed for various HTS
\cite{Zhao95,Khasanov03,Khasanov03a,Khasanov04,Khasanov04a,Khasanov06,Zhao97,
Zhao98,Hofer00,Zhao01}. For all the samples the oxygen isotope shift
$\Delta\lambda_{ab}(0)/\lambda_{ab}(0)$ decreases by more than a
factor of two with increasing the magnetic field from 0.2~T to 0.6~T
(see Table~\ref{Table:OIE-results}). This implies that
$^{16}\lambda_{ab}(0)$ increases stronger with field than
$^{18}\lambda_{ab}(0)$,  as expected for $^{16}m_{ab}^\ast<\
^{18}m_{ab}^\ast$ (see above).

\begin{figure}[htb]
%\centering
\includegraphics[width=1.05
\linewidth]{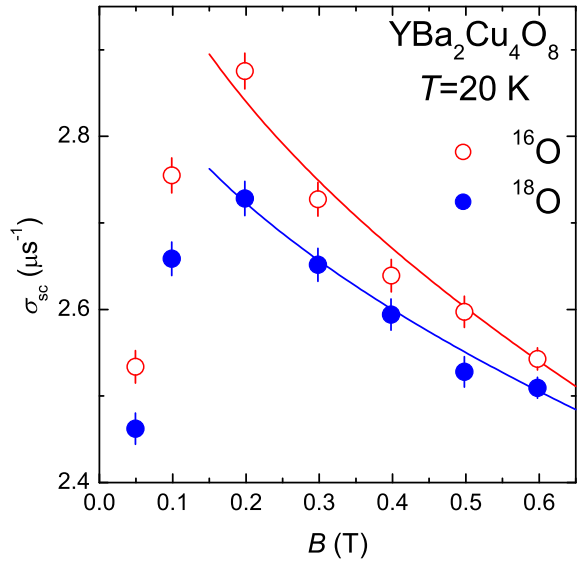}
%\vspace{-0.8cm}
%
\caption{(Color online) Magnetic field dependence of the $\mu$SR
depolarization rate $\sigma_{sc}$ for $^{16}$O ($\circ$) and
$^{18}$O ($\bullet$) substituted YBa$_2$Cu$_4$O$_8$ samples at
$T=20$~K. Solid curves represent fits of the data using
Eq.~(\ref{eq:lambda_vs_H}) (see text for details). }
 \label{figure:field_dependence}
\end{figure}

I order to study the magnetic field dependence of $\lambda_{ab}$
in more detail, $\sigma_{sc}$ was measured as a function of $H$ at
$T=20$~K for the $^{16}$O and $^{18}$O substituted
YBa$_2$Cu$_4$O$_8$ samples (see
Fig.~\ref{figure:field_dependence}). For both samples
$\sigma_{sc}$ first increases, goes through a pronounced maximum
around $\mu_0H\sim0.2$~T and than decreases.
Note, that for conventional $s-$wave superconductors, where
$\lambda$ is field independent, $\sigma_{sc}$ increases with
increasing magnetic field up to $H\simeq 2H_{c1}$ and then stays
constant for fields much smaller than $H_{c2}$ \cite{Brandt88}
($H_{c1}$ and $H_{c2}$ are the first and the second critical
fields, respectively). For $d-$wave superconductors, however,
$\lambda$ depends on the field. This leads to a decrease of
$\sigma_{sc}(H)$ for magnetic fields higher that $2H_{c1}$.
Such behavior is generally observed for various HTS
\cite{Zimmermann95,Sonier00}. The solid lines represent results of
the fits by means of the relation
\begin{equation}
\frac{\lambda^{-2}_{ab}(B)}{\lambda^{-2}_{ab}(B=0)}=
\frac{\sigma_{sc}(B)}{\sigma_{sc}(B=0)}= 1-K\cdot \sqrt{B},
 \label{eq:lambda_vs_H}
\end{equation}
which takes into account the nonlinear correction to $\lambda_{ab}$
\cite{Won01,Vekhter99}. Here $K$ is the parameter depending on the
strength of the nonlinear effect. In the analysis only the points
above 0.1~T were considered. The magnetic field inside the sample
was assumed to be equal to the external magnetic field $B\simeq
\mu_0H_{ext}$. The fits yield: $^{16}\sigma_{sc}(B=0)=3.25(3)$~$\mu
s^{-1}$, $^{16}K$=0.282(5)~T$^{-1/2}$, and
$^{18}\sigma_{sc}(B=0)=3.02(3)$~$\mu s^{-1}$,
$^{18}K$=0.220(5)~T$^{-1/2}$ for the $^{16}$O and $^{18}$O
substituted samples, respectively. The larger $^{16}\sigma_{sc}(B)$
and its stronger field decrease in comparison with
$^{18}\sigma_{sc}(B)$ is consistent with the finding
$^{18}m^\ast_{ab}>$$^{16}m^\ast_{ab}$
\cite{Zhao95,Khasanov03,Khasanov03a,Khasanov04,Khasanov04a,Khasanov06,Zhao97,
Zhao98,Hofer00,Zhao01} and with the statement that the nonlinear and
the nonlocal corrections to $\lambda_{ab}$ decrease with increasing
in-plane supercarrier mass $m_{ab}^\ast$.
One should mention, however, that the measurements were performed
in magnetic fields well below $H_{c2}$ ($0.0006\leq H/H_{c2}\leq
0.008$), so that it is difficult to draw any firm conclusion about
the precise field dependence of $\lambda_{ab}$. In particular, the
stronger field dependence of $^{16}\sigma_{sc}$ suggests that at
some characteristic field the sign of the OIE on $\lambda_{ab}$
should change. Further investigations at higher fields are needed
to clarify this point.

All the above observed features are consistent with the picture
where the mass of the charge carriers $m^\ast_{ab}$ depends on the
oxygen isotope mass. It stems from the isotope dependence of the
in-plane magnetic penetration depth $\lambda_{ab}$, that was found
to be larger for the samples with the heavier oxygen isotope
($^{18}\lambda_{ab}>$$^{16}\lambda_{ab}$). An additional
confirmation comes from the stronger magnetic field dependence of
$\lambda_{ab}$ in the $^{16}$O substituted samples. The later one is
determined by the fact that both the nonlinear and the nonlocal
corrections to $\lambda_{ab}$, arising from the presence of nodes in
the superconducting gap, decrease with increasing $m_{ab}^\ast$.

In conclusion, the magnetic field dependence of the OIE on the
in-plane magnetic field penetration depth $\lambda_{ab}$ was
studied in YBa$_2$Cu$_4$O$_8$,
Y$_{0.8}$Pr$_{0.2}$Ba$_2$Cu$_3$O$_{7-\delta}$, and
Y$_{0.7}$Pr$_{0.3}$Ba$_2$Cu$_3$O$_{7-\delta}$ hole-doped HTS. As a
result $\lambda_{ab}$ for the $^{16}$O substituted samples
increases stronger with magnetic field than for the $^{18}$O
substituted ones. The OIE on $\lambda_{ab}$ was found to decrease
from $\Delta\lambda_{ab}(0)/\lambda_{ab}(0)=2.2(6)$\%, 2.4(4)\%,
and 2.1(6)\% at $\mu_0H=0.2$~T to 0.9(6)\%, 1.0(4)\%, and 0.9(6)\%
at $\mu_0H=0.6$~T for YBa$_2$Cu$_4$O$_8$,
Y$_{0.8}$Pr$_{0.2}$Ba$_2$Cu$_3$O$_{7-\delta}$, and
Y$_{0.7}$Pr$_{0.3}$Ba$_2$Cu$_3$O$_{7-\delta}$, respectively. Both
of the above mentioned effects can be explained by the isotope
dependence of the in-plane charge carrier mass $m^\ast_{ab}$.

This work was partly performed at the Swiss Muon Source (S$\mu$S),
Paul Scherrer Institute (PSI, Switzerland). The authors are grateful
to A.~Amato and D.~Herlach for assistance during the $\mu$SR
measurements. This work was supported by the Swiss National Science
Foundation in part by the EU Project CoMePhS, and by the
K.~Alex~M\"uller Foundation.

\end{document}